\documentclass[aps,prl,superscriptaddress,showpacs,twocolumn,floatfix,amsmath,amssymb,10pt]{revtex4-2}
\usepackage{hyperref} 
\usepackage{graphicx} 
\usepackage{xcolor}
\usepackage[version=4]{mhchem}
\usepackage{graphicx}

\def\micro{\mu}

\def\gtorder{\mathrel{\raise.3ex\hbox{$>$}\mkern-14mu
 \lower0.6ex\hbox{$\sim$}}}
\def\ltorder{\mathrel{\raise.3ex\hbox{$<$}\mkern-14mu
 \lower0.6ex\hbox{$\sim$}}}
\def\etal{\textit{et al.}}

\begin{document}
\title{Precise measurements of the $^3$H/$^3$He charge form factor ratio and their impact on the $^3$H-$^3$He charge radius difference}

\author{L.~Kurbany} \affiliation{University of New Hampshire, Durham, New Hampshire 03824, USA}

\author{S.~Li} \affiliation{University of New Hampshire, Durham, New Hampshire 03824, USA} \affiliation{Lawrence Berkeley National Laboratory, Berkeley, California 94720, USA} 

\author{S.~N.~Santiesteban}
\affiliation{University of New Hampshire, Durham, New Hampshire 03824, USA} 

\author{J.~Arrington}
\email{JArrington@lbl.gov}
\affiliation{Lawrence Berkeley National Laboratory, Berkeley, California 94720, USA}
\affiliation{Argonne National Laboratory, Lemont, Illinois 60439, USA} 

\author{D.~Abrams} \affiliation{University of Virginia, Charlottesville, Virginia 22904, USA} 


\author{S.~Alsalmi } \affiliation{Kent State University, Kent, Ohio 44240, USA} 
\affiliation{King Saud University, Riyadh 11451, Kingdom of Saudi Arabia}

\author{D.~Androic} \affiliation{University of Zagreb, Zagreb, Croatia} 

\author{K.~Aniol} \affiliation{California State University, Los Angeles, California 90032, USA} 



\author{T.~Averett} \affiliation{William and Mary, Williamsburg, Virginia 23185, USA} 

\author{C.~Ayerbe~Gayoso} \affiliation{William and Mary, Williamsburg, Virginia 23185, USA} 


% ...


\date{\today}

\begin{abstract}

\textbf{Background:} Nuclear form factor measurements at low momentum transfers can be used to extract the nuclear charge radius. These can be compared to \textit{ab initio} nuclear structure calculations, with the difference between $^3$H and $^3$He charge radii providing information on the spatial distribution of protons vs neutrons in these nuclei. Extractions of the $^3$H - $^3$He charge radius difference are limited by the modest amount of data for tritium elastic scattering at low momentum transfer.  

\textbf{Purpose:} The goal of the experiment is to make a precise measurement of the $^3$H/$^3$He form factor ratio at low momentum transfer. The ratio is then combined with the absolute $^3$He measurements to provide a $^3$H form factor measurement with improved precision compared to world data. This will provide an improved determination of the absolute normalization corrections of world data sets, providing improved precision for the low-$Q^2$ measurements that are most sensitive to the radius.

\textbf{Methods:} We measure elastic scattering from nearly identical $^3$H and $^3$He targets. This allows for an extraction of the $^3$H/$^3$He form factor ratio, with significantly reduced systematic uncertainty, as identical systematic uncertainties will cancel in the target ratio. 

\textbf{Results:} We extract the elastic cross sections and the $^3$H/$^3$He ratio at a single value of momentum transfer. We use this to provide a high-precision measurement of the $^3$H charge form factor at $Q^2=3.03$~fm$^{-2}$ and evaluate the impact of this point on a global fit of previous form factor extractions. We examine both the impact on the normalizations of the other data sets and the fit parameters, including the charge radius and its uncertainty. 

\textbf{Conclusions:} Our sensitivity study shows that the new precision data points constrain the normalization of previous data sets, reducing the uncertainties by up to 12\% and yielding a 4\% improvement in the radius. Including projected data based on an approved experiment yields improvements in the normalization factors of up to 20\% and a 27\% improvement in the $^3$H RMS charge radius extraction, allowing  for an improved extraction of the $^3$H-$^3$He radius difference from a direct fit to the form factor ratios over a range in $Q^2$.

\end{abstract}

\maketitle

\section{Introduction}

Elastic scattering from nuclei provides access to the nuclear electromagnetic form factors. For spin-1/2 nuclei, such as $^3$H and $^3$He, there are two quantities that enter into the elastic cross section: the charge ($F_c(Q^2)$) and magnetic ($F_m(Q^2)$) form factors. These can be extracted by measuring the elastic scattering cross section as a function of scattering angle, $\theta$, at fixed four-momentum transfer squared, $Q^2$~\cite{Rosenbluth:1950yq}. The form factors depend only on $Q^2$ and, in the non-relativistic limit, are the Fourier transforms of the nuclear charge and magnetization densities~\cite{Miller:2009sg}. We use the convention in which both form factors are normalized to unity at $Q^2 = 0$, $F_c(0) = F_m(0) = 1$, with the nuclear charge $Z$ and magnetic moment $\mu$ appearing explicitly in the cross section. The slope of the form factors $F_c(Q^2)$ ($F_m(Q^2)$) as $Q^2 \to 0$ is proportional to the mean-square (MS) charge (magnetic) radius for the nucleus.

Comparisons of the $^3$H and $^3$He charge radii can be used as a precise test of \textit{ab initio} nuclear structure calculations. In addition, atomic physics measurements provide extremely precise measurements of the difference in charge radius between different hydrogen isotopes and the difference in helium isotopes~\cite{Jentschura:2011mdl, Mueller:2007dhq, Mohr:2024kco}. An improved extraction of the $^3$H-$^3$He charge radius difference will provide an improved measurement of the difference between these isotope chains. There are many measurements of elastic e-$^3$He scattering at low $Q^2$ values but fewer of elastic e-$^3$H scattering, typically with much larger uncertainties, due to the difficulties involved in making a target from a non-stable nucleus~\cite{collard65, Beck:1989bi, AMROUN1994596}. Thus, the difference between the $^3$H and $^3$He charge radii is limited by our knowledge of the tritium elastic form factors. 

The form factors are extracted from scattering cross section measurements that quantify the reaction probability for elastic electron-nucleus scattering. For $^3$H and $^3$He, there are two form factors that parameterize the cross section deviation relative to scattering from a structureless spin-1/2 particle. The elastic cross section deviates from the point-like Mott cross section by a factor equal to:
\begin{equation}
\frac{1}{\eta}  \bigg[\frac{Q^2}{q^2}F^2_{c}(q)+\frac{\mu^2 Q^2}{2M^2}\left(\frac{1}{2}\frac{Q^2}{q^2}+ \tan^2 \left(\frac{\theta}2\right)\right)F^2_{m}(q)\bigg]
  \label{eq:born cross-section}
\end{equation}
where $q$ is the three-momentum and $Q^2$ is the four-momentum transfer squared of the virtual photon, $\eta = 1+ Q^2 / (4M^2)$, where $M$ is the mass of the nucleus, and $\mu$ is its magnetic moment.

In the non-relativistic limit, the form factor can be expanded in terms of $Q^2$ as
\begin{equation}
  F_c(Q^2)\approx 1 - \frac{Q^2}{6} \langle r^2\rangle +   \frac{Q^4}{120} \langle r^4\rangle + ...,
  \label{eq:ffexpansion}
\end{equation}
where $\langle r^2\rangle$ is the mean-square (MS) charge radius of the nucleus. While there are corrections in the relativistic case, the conventional definition for the MS charge radius is taken to be the slope of the form factor at $Q^2=0$.
\begin{equation}
    \langle r^2\rangle = - 6 \left.\frac{dF_c(Q^2)}{dQ^2}\right|_{Q^2=0} \approx \frac{6 (1 - F_c(Q^2))}{Q^2}~\mbox{for}~{Q^2 \approx 0}
    \label{eq:radius}
\end{equation}

Because the form factors of both nuclei are normalized to unity at $Q^2=0$, the ratio of the $^3$H and $^3$He charge form factors at low $Q^2$ is directly sensitive to the difference between their mean-square charge radii. Keeping only the leading term of Eq.~\ref{eq:ffexpansion},
\begin{equation}
  \frac{F_c(^3\mathrm{H})}{F_c(^3\mathrm{He})} \approx
  \frac{1 - Q^2 \langle r^2\rangle_{^3\mathrm{H}}/6}
       {1 - Q^2 \langle r^2\rangle_{^3\mathrm{He}}/6}
  \approx 1 - \frac{Q^2}{6}\left(\langle r^2\rangle_{^3\mathrm{H}}
  - \langle r^2\rangle_{^3\mathrm{He}}\right)
  \label{eq:ratio}
\end{equation}
Many experimental uncertainties cancel in the ratio, so a precise measurement of the form factor ratio at low $Q^2$ provides a direct constraint on the $^3$H--$^3$He radius difference. This approximation is only valid for $Q^2 \langle r^2\rangle \ll 6$. At larger $Q^2$, the higher-order terms in Eq.~\ref{eq:ffexpansion} become significant, and a fit including the $\langle r^4\rangle$ and higher terms is required.

Jefferson Lab (JLab) experiment E12-11-112~\cite{e1211112} was conducted in Hall A during the fall of 2018, with data taken at three different beam energies. The majority of the experiment measured quasielastic (QE) scattering from $^3$H and $^3$He, with beam energies of approximately 2.2 and 4.3~GeV~\cite{Arrington:2023hht, Li:2022fhh, Santiesteban:2023rsh, Li:2024rzf}. A small subset of data was taken with a beam energy of 1.171~GeV, focused on a precise measurement of the ratio of nuclear elastic scattering from $^3$H and $^3$He. Taking the ratio of elastic scattering for $^3$H and $^3$He allows for improved systematic uncertainty, as many contributions will be identical or nearly identical between the two targets. We can then use this precise measurement of the ratio, combined with the more extensive existing measurements on $^3$He, to improve our understanding of the tritium charge radius and the $^3$H-$^3$He difference. This was originally proposed as JLab experiment E12-14-009~\cite{e1214009} with the goal of measuring small-angle elastic scattering for $^3$H and $^3$He, covering $1.2 < Q^2 < 2.3$~fm$^{-2}$. In this region, the sensitivity to the charge radius is maximized, allowing for a significant improvement in the measurement of the $^3$H-$^3$He charge radius difference. In addition, the low $Q^2$ values are most sensitive to $F_c$, with minimal contributions from $F_m$, allowing the cross section ratio to be used to make a precise extraction of the $F_c$ ratio. During the E12-11-112 experiment, we were able to take a single point at $\theta=17$~degrees, allowing for the cross section ratio at $Q^2=3.03$~fm$^{-2}$. This leads to a larger $F_m$ contribution and provides reduced sensitivity to the charge radius, but still provides a new, high-precision data point that better constrains fits of the charge form factor.

\section{Experimental Setup}

For this measurement, a 1.171~GeV electron beam impinged on $^1$H, $^2$H, $^3$H, and $^3$He targets in Hall A. Scattered electrons were detected in the left high-resolution spectrometer (LHRS). The spectrometer central angle was set to $17^\circ$ with the central momentum tuned to 1.128~GeV to measure nuclear elastic scattering from these targets. The Continuous Electron Beam Accelerator Facility (CEBAF) has two superconducting RF linacs connected by recirculator arcs that can each provide up to 1.1~GeV to the accelerated electrons~\cite{Leemann:2001dg, Alcorn:2004sb}. The arcs allow for a total of 4 passes through the LINACS before the electrons are delivered to Hall A. For this experiment, CEBAF was run in a non-standard mode to provide the lower electron energies needed to access the low $Q^2$ values required. The machine was set up in a single-LINAC configuration, providing a beam energy of 1.171~GeV. Upon entering Hall~A, the electron beam passed through a series of beamline components and monitoring systems~\cite{Alcorn:2004sb}, including beam charge monitors (BCMs) for current measurement, rastering systems to mitigate localized target heating, and beam position monitors (BPMs) to determine the beam position prior to its interaction with the target~\cite{Alcorn:2004sb}.

The experiment used a novel high-pressure sealed-cell target design, described in detail in Refs.~\cite{target_NIM, Arrington:2023hht}. The targets were 25~cm long, with four cells designed to hold 10 atmospheres of $^1$H, $^2$H, or $^3$H, and 20 atmospheres of $^3$He. The $^3$H and $^3$He cells used for this measurement had total thicknesses of 84.8$\pm$0.4 and 53.2$\pm$0.5~mg/cm$^2$, respectively, as discussed in more detail in Refs.~\cite{Li:2022fhh, Abrams:2021usd}. Energy deposited by the electron beam as it passes through the target causes heating of the target gas, leading to local density changes in the target. This effect requires what is typically referred to as a ``target boiling'' correction, which was analyzed for these targets in Ref.~\cite{Santiesteban:2018qwi}. Because the average beam current for the elastic scattering measurement was only 5~$\micro$A, the correction is very small: roughly 3\% for $^3$H and 2\% for $^3$He~\cite{kurbani_thesis}. 

The LHRS uses a QQDQ design where the quadrupoles are responsible for electron focusing and the dipole disperses the electrons, facilitating momentum measurement~\cite{Alcorn:2004sb}. The scattered electrons coming through the final quadrupole of the LHRS then pass through a detector package containing two vertical drift chambers (VDCs), two scintillator planes (S$_0$ and S$_2$), a gas Cerenkov detector, and a calorimeter. The VDCs provide precise position information, allowing for a reconstruction of the position and angle of the track in the detectors. 
The reconstructed track, combined with a detailed model of the spectrometer optics, allows the particle kinematics at the target to be reconstructed with a momentum resolution better than $2\times10^{-4}$ and an in-plane scattering angle resolution of 2~mrad over a range of central momentum settings from 0.3~GeV to 4.0~GeV~\cite{Ou:2019ing}. 

The scattered electrons then traverse two scintillator planes, S$_0$ and S$_2$, which are used to form triggers for the Data Acquisition (DAQ) system and provide precise event timing information. The Cerenkov detector is used in the trigger and in the final analysis to separate electrons from hadrons (mainly pions) based on their very different thresholds for Cerenkov radiation emission. Finally, the scattered electrons traverse the electromagnetic calorimeters, which can separate electrons from hadrons based on the measured energy deposition. The overall detector package is described in Refs.~\cite{Alcorn:2004sb, kurbani_thesis}.

\section{Data Analysis}

The scattering cross section is extracted by counting the number of elastically scattered electrons, applying corrections for detector efficiency, spectrometer acceptance, and other losses, and then normalizing the yield to the integrated luminosity. Since the observed cross section encompasses contributions from multiple scattering, radiative energy loss, and QED radiative corrections, a series of corrections must be applied to recover the underlying Born cross section. Corrections accounting for detector performance, event selection criteria, and data acquisition live time are incorporated as overall normalization factors, and are absorbed into the definition of the luminosity-normalized yield:
\begin{equation}
Y_{norm}^{dat}=\frac{N_{el} \cdot f_{ps}}{N_{in}\cdot\rho\cdot\Delta Z\cdot L_C\cdot\varepsilon_{eff}}
\label{eq:yield}
\end{equation}
where $N_{el}$ is the number of elastically scattered electrons detected after the application of selection cuts, $f_{ps}$ is the prescale factor for the production trigger, $N_{in}$ is the total number of electrons incident on the target, $\rho$ is the target density in g/cm$^3$, $\Delta Z$ is the effective target length in cm, $L_C$ is the computer live time, and $\varepsilon_{eff}$ is the product of all detector and cut efficiencies, including trigger, tracking, and particle identification (PID) efficiencies.

Corrections associated with spectrometer acceptance, radiative corrections, and multiple scattering are evaluated through a detailed Monte Carlo simulation of the experiment. This approach allows the corrections to be applied consistently within a unified framework, without requiring the assumption that they factorize from one another. The simulation yields a luminosity-normalized simulated yield, $Y_{\mathrm{norm}}^{\mathrm{sim}}$, based on the cross section model employed in the simulation. The ratio of the measured yield to the simulated yield is then computed and applied as a multiplicative correction to the underlying cross section model. A comprehensive description of the analysis procedure is provided in Ref.~\cite{kurbani_thesis}, with a summary of the applied cuts, corrections, and efficiencies presented in the following sections. Details of the cuts that were applied to isolate elastic events are described below.

\begin{itemize}
\item \textbf{Acceptance cuts}: We apply a series of cuts to the reconstructed target quantities to limit the events to the region where the acceptance of the LHRS is well understood. We apply cuts on the out of plane angle $\theta_{tg}$, in plane angle $\phi_{tg}$, and the momentum fraction $\delta=(P-P_0)/P_0$, where $P$ is the reconstructed electron momentum and $P_0$ is the central momentum setting of the LHRS. The cuts used for the cross section extraction (`tight cut') and those used to study the cut dependence (`loose cut') are given in Table~\ref{tab:acceptance-cuts}. In addition, a target length cut is applied to reduce the contribution of events coming from the aluminum endcaps (entrance and exit windows) of the target cell. We subtract the residual endcap events using measurements on an empty target cell, described below.

\begin{table}[tbh]
\begin{center}
\begin{tabular}{c|cc|cc}
\hline
Cut quantity               &\multicolumn{2}{c}~~Tight cut~~~~&\multicolumn{2}{|c}~~Loose cut~~~~\\ \hline 
                           &~~Min & Max &~~Min & Max \\    \hline 
    $\theta_{tg}$ [mrad]   & -35 & 35  & -50 & 50  \\
    $\phi_{tg}$ [mrad]     & -25 & 25  & -35 & 35  \\
    $\delta$ [\%]          & -3.5& 3.5 & -5.0& 5.0  \\
    $z_{tg}$ [cm]          & -8  & 8   & -8  & 8   \\
\hline
\end{tabular}
\caption{Cuts on the reconstructed target variables. The tight cuts are applied to study the physics variables and loose cuts are used to estimate the efficiencies.}
\label{tab:acceptance-cuts}
\end{center}
\end{table}

\item \textbf{Particle identification (PID) Cuts}: We apply cuts on the Cerenkov and calorimeter detector signals to isolate scattered electrons from other negatively-charged particles, predominantly pions. Pions typically do not trigger the Cerenkov detector but produce a small signal in the calorimeter. We require that the energy detected in the calorimeter is at least 70\% of the energy of the incoming particle, assuming it's an electron, which yields a 99.4\% efficiency for electrons. We apply a threshold on the signal from the Cerenkov detector that is 99.9\% efficient for electrons but yields significant pion rejection. Because the initial pion fraction is low for these kinematics, the final pion contamination is negligible, and the combined efficiency is 99.3\%.

\item \textbf{Trigger efficiency:} The trigger efficiency is defined as the fraction of good electron events that successfully generate a trigger in the HRS. It is associated with inefficiencies of the detectors used to form the trigger: the hodoscopes and the gas Cerenkov detector. The main trigger requires signals from both scintillator planes (S$_0$ and S$_2$) and the Cerenkov. To determine the efficiency of the discriminator cuts on these detectors used in the trigger, we have two efficiency triggers: one where we require only one of the scintillator planes plus the Cerenkov  and one where we require signals in both scintillator planes but not the Cerenkov. We take a sample of these events during each run to determine the efficiency of the main trigger. This efficiency is evaluated on a run-by-run basis and consistently exceeds $99\%$ with negligible uncertainty.

\item \textbf{Tracking efficiency:} There is a small probability that a good electron event will not yield a track in the VDCs. In addition, there can be two particles passing through the VDCs close enough together that their hits overlap in time, complicating track reconstruction. To avoid selecting a track that does not correspond to the event that provided the trigger, we use only events where one and only one track is identified. A correction is applied to account for valid events lost due to multiple or no tracks being found.

\item \textbf{Live-time correction $L_C$ and prescale factor $f_{ps}$:}
The live-time correction addresses events that do not form a trigger due to electronic dead time in the DAQ system or that are not read out because the data acquisition is busy when the trigger arrives. To keep the computer live time high, the production trigger was prescaled by a factor of $f_{ps}=25$ (only every 25th trigger was read out). The computer live time was determined from the ratio of the number of events read out by the DAQ to the total number of triggers recorded by the scalers, corrected for the prescale factor. The live time was 92\% for both the $^3$H and $^3$He runs, with a statistical uncertainty of about 0.01\%. For the rates of this experiment, the electronic dead time is negligible, so no correction is applied. Because the $^3$H and $^3$He data were taken under nearly identical rate conditions, any residual rate-dependent effect largely cancels in the cross-section ratio.

\item \textbf{Endcap contribution:}
In addition to measuring scattering from the gas in the target cells, the spectrometer also detects events where the electron scatters in the target's aluminum endcaps. The cut on $z_{tg}$, the distance from the center of the target along the beam direction, significantly suppresses these events, and the residual contamination is subtracted using measurements on an empty target cell of the same geometry as the gas cells, which provides a clean measurement of scattering from the endcaps. The empty-cell data are subtracted after scaling the entrance- and exit-window yields to match the endcap peaks of the $^3$H or $^3$He target measurement, accounting for small differences in window thickness and shape between the individually machined cells. The size of this subtraction is approximately 1.5\% for $^3$H and 1.3\% for $^3$He within the $-8 < z_{tg} < 8$~cm acceptance. The associated uncertainty, estimated from the spread of the extracted endcap scaling factors, is 0.1\% on the absolute cross sections and 0.05\% on the $^3$H/$^3$He ratio. 

\end{itemize}

The beam charge was determined from the Hall~A beam current monitors (BCMs). The BCMs are calibrated against the absolute Unser monitor~\cite{Alcorn:2004sb, kurbani_thesis}, in special runs alternating between zero current and various periods with beam at different currents~\cite{kurbani_thesis}. The charge uncertainty is relatively large (3\%) because these data were taken at an average beam current of approximately 5~$\mu$A, while the BCMs are optimized for higher currents; the other measurements from this run period~\cite{cruz-torres19, cruz-torres20, Li:2022fhh, Abrams:2021usd, Santiesteban:2023rsh, Abrams:2024las, Li:2024rzf} used currents of roughly 20~$\mu$A. This uncertainty is dominated by the BCM offset and gain calibration, which is common to both targets. Since the $^3$H and $^3$He data were taken at nearly identical beam currents, it largely cancels in the ratio, contributing only 0.2\% to the $^3$H/$^3$He cross-section ratio.

While the `target boiling' correction was applied as a modification to the nominal target density and scattering from the aluminum endcaps was subtracted, there are two additional target contamination effects that need to be addressed. Tritium is a radioactive isotope that decays into helium-3 with a half life of approximately 15 years. Thus, the $^3$H target has a small $^3$He contamination that increased over time during the course of the experiment. Based on the decay lifetime and the time since the $^3$H target was filled, we determined that $(1.4 \pm 0.1)\%$ of the $^3$H had decayed to $^3$He at the time of the data taking, and we used the measured $^3$He cross section to apply a correction to the $^3$H yield.

It was also determined that the $^3$H cell used during this part of the measurement contained a small amount of  $^1$H, based on the observation of a small peak for elastic e-$^1$H scattering. It was concluded that the target used during the experiment had a small amount of water as contamination. When the target was cooled, the water froze to the cell walls and thus did not interact with the beam. However, the presence of water in the cell allowed for the exchange of $^3$H and $^1$H atoms between the water and the tritium gas, resulting in the small elastic peak of e-$^1$ H and reducing the amount of $^3$H available for scattering. This was quantified by measurements at the elastic electron-proton peak~\cite{Li:2022fhh, Santiesteban:2018qwi}, leading to a (4.12$\pm$0.20)\% reduction in the total amount of $^3$H available for scattering. Although the electron beam also scatters from the $^1$H nuclei in the gas, this is kinematically limited to $x \le 1$, while the elastic scattering from $^3$H is measured at $x \approx 3$, so contamination of $^1$H does not affect the observed yield of nuclear elastic scattering.

\section{Extraction of the cross section and charge form factor}

The SIMC Monte Carlo package~\cite{Gaskell:2001, Mohring:2002cpu} was used to simulate elastic electron scattering from $^{3}$H, $^{3}$He, and $^{1}$H targets. The code was modified to update the elastic e-p scattering option to simulate scattering from any nucleus, given its mass and a parameterization of its elastic cross section. SIMC generates events with a weight determined by the nuclear elastic cross sections, using the form factor parameterization of Ref.~\cite{AMROUN1994596} for $^3$H and $^3$He.  It simulates energy loss, multiple scattering, Coulomb corrections, and radiative corrections~\cite{Ent:2001hm}, as well as the acceptance of the spectrometer. Rather than correcting the yield for each of these effects individually, we use the simulation to determine the predicted event yield in the real experiment, given a specific value for the cross section at the central kinematics. Thus, the cross section we extract from the data is taken to be:
\begin{equation}
d\sigma / d\Omega = Y^{dat}_{norm} \frac{(d\sigma / d\Omega)^{sim}}{Y^{sim}_{norm}}
\label{eq:xsecextract}
\end{equation}
where the extracted and simulated values for $d\sigma / d\Omega$ correspond to the Born cross section. Coulomb distortion of the incoming and outgoing electron waves is included in the cross section model using the effective momentum approximation (EMA) that has been tuned against \textit{ab initio} calculations based on the prescription of Ref.~\cite{Aste:2007hh}. 

\begin{figure}[htb]
\includegraphics[width=0.85\columnwidth, trim={0 0 0 0}, clip]{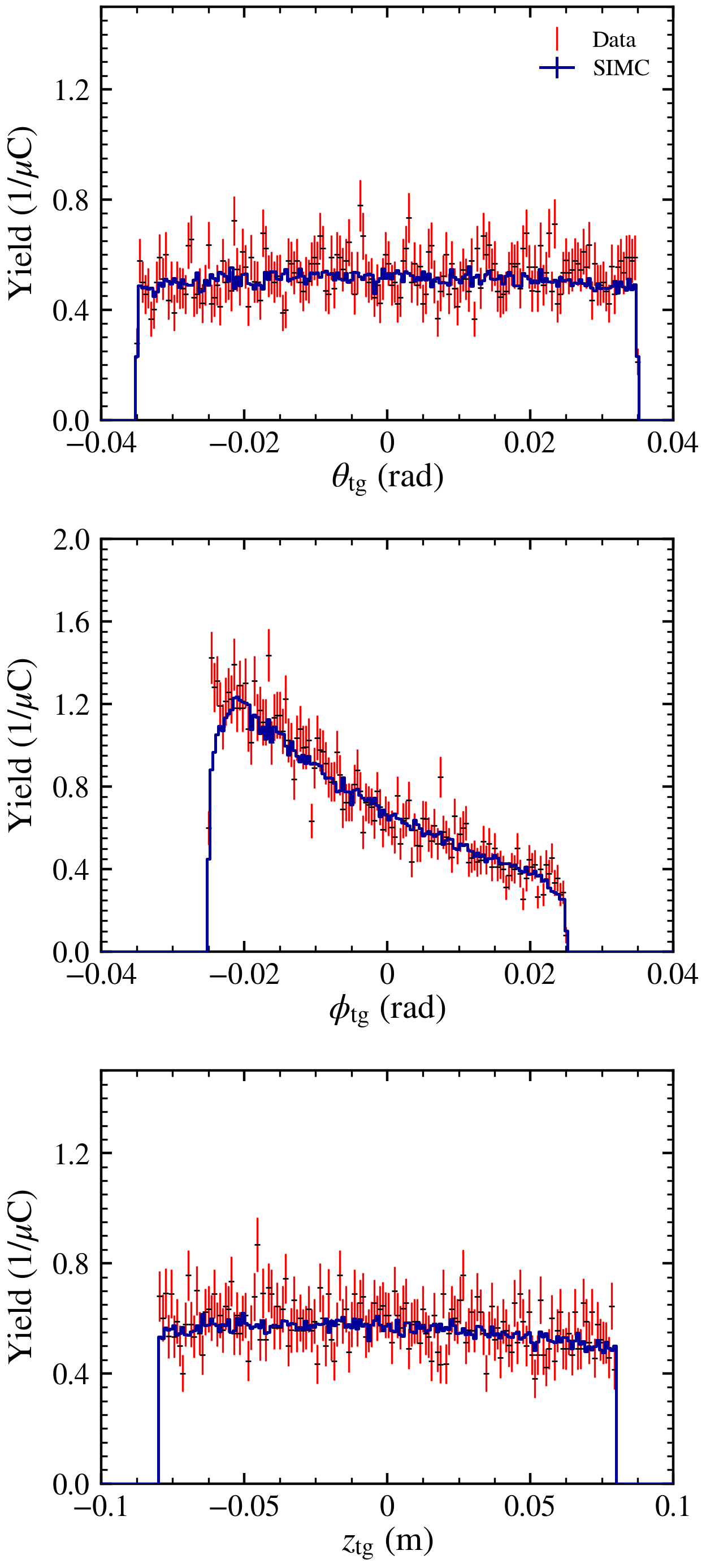}
 \caption{Comparison of reconstructed data and SIMC distributions for the $^3$H target.}
 \label{fig:simcfit}
\end{figure}

\begin{figure}[htb]
\includegraphics[width=0.9\columnwidth]{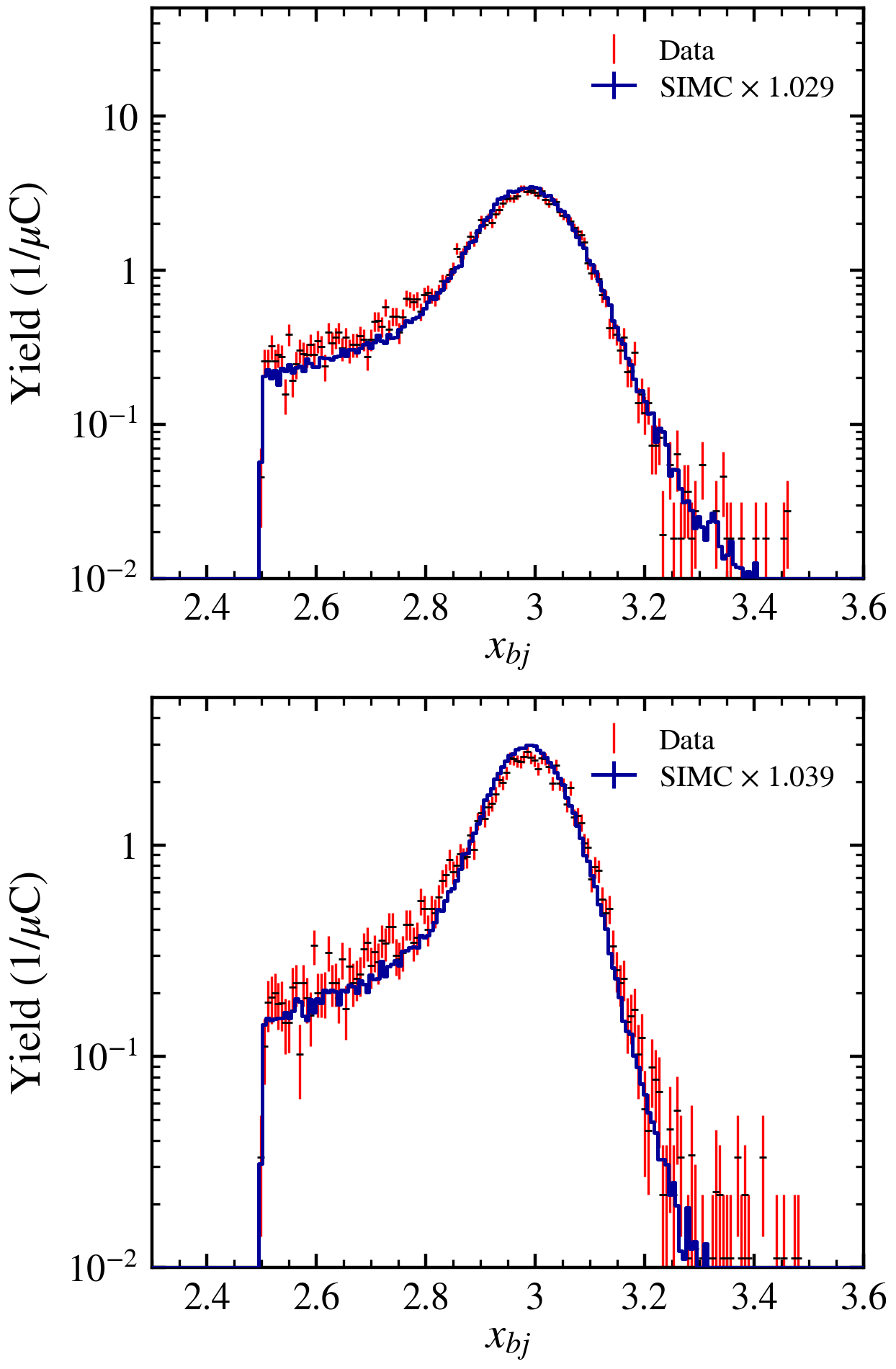}
 \caption{Comparison of data and SIMC elastic peaks.}
 \label{fig:data-simc}
\end{figure}

Figure~\ref{fig:simcfit} compares the simulated and measured distributions as a function of $\theta_{tg}$, $\phi_{tg}$, and $z_{tg}$ for the $^3$H data with the standard analysis cuts applied. The distributions are in excellent agreement, with equally good agreement for the $^3$He data. Figure~\ref{fig:data-simc} shows the simulated and measured distributions in the vicinity of the nuclear elastic peak and includes the $x$ cut applied to remove inelastic events. While both energy loss and radiative effects are simulated on an event-by-event basis, we apply a correction for the most probable energy loss to the data and simulation so that the elastic peak is centered close to the expected value of $x = M_A/M_p$. The incident electron beam loses approximately 0.3~MeV passing through the target entrance window and gas, while scattered electrons also lose around 1 MeV due to ionization as they travel through the target wall, scattering chamber exit window, and spectrometer entrance window. 

Table~\ref{tab:systematics} shows the systematic uncertainties in the extraction of the elastic cross sections and in the $^3$H/$^3$He cross section ratio. For the absolute cross sections, the dominant sources of uncertainty are associated with the measured beam charge, the spectrometer acceptance, and the radiative corrections. The acceptance uncertainty is also larger than that typical for other experiments using the LHRS because we limit ourselves to the central 16~cm of the target by applying a cut on the reconstructed target length. A more complete description of the cross section uncertainties is provided in~\cite{kurbani_thesis}.

\begin{table}[htb]
\centering
\caption{Summary of the systematic uncertainties on elastic cross sections and the $^3$H/$^3$He cross section ratios.}\label{tab:systematics}
\begin{tabular}{|l|c|c|}
    \hline
    \textbf{Source}     & \textbf{$d\sigma/\sigma$}         &~~~\textbf{$dR/R$}~~~ \\ \hline
    Charge              & 3.0\%                             & 0.2\%     \\ \hline
    Target thickness    & 0.4\%($^3$H), 1.0\%($^3$He)       & 1.06\%    \\ \hline
    Target boiling      & 0.5\%                             & 0.1\%     \\ \hline
    Hydrogen contam.    & 0.2\% ($^3$H only)                & 0.2\%     \\ \hline
    $^3$He contam.      & 0.1\% ($^3$H only)                & 0.1\%     \\ \hline  
    Endcap contam       & 0.1\%                             & 0.05\%    \\ \hline
    Trigger/PID cuts    & 0.1\%                             & 0.05\%    \\ \hline
    Tracking Eff        & 0.2\%                             & 0.05\%    \\ \hline
    Acceptance          & 2.0\%                             & 0.2\%     \\ \hline
    LHRS $\theta$ offset & 0.9\%                             & 0.1\%    \\ \hline
    Beam energy offset  & 0.3\%                             & 0.05\%    \\ \hline
    Rad Corr            & 1.5\%                             & 0.3\%     \\ \hline \hline
    \textbf{Total}      & 4.08\% ($^3$H), 4.18\% ($^3$He)   & 1.17\%    \\ \hline
      \end{tabular}
    \end{table}

Many sources of uncertainty are common to the measurements on the $^3$H and $^3$He so their contributions to the uncertainty in the cross section ratio are significantly reduced. Because the design of the target cells is identical and the measurements were taken at nearly identical beam currents, the uncertainties in the acceptance and the beam charge measurement contribute minimally to the ratio. The main sources of uncertainty on the ratio are those associated with the target thicknesses, density fluctuations, and hydrogen contamination in the $^3$H target~\cite{Santiesteban:2018qwi, Arrington:2023hht}, yielding a total systematic uncertainty of 1.17\% on the yield ratio. The statistical uncertainties are of order 1\% for $^3$H and $^3$He, yielding a statistical uncertainty of 1.41\% on the yield ratio. Combining the statistical and systematic contributions, we obtain total uncertainties of 4.2\% (4.3\%) for the $^3$H ($^3$He) cross section and 1.84\% for the cross section ratio.

After applying the acceptance cuts to both the experimental data and the SIMC simulation yield and correcting the experimental yield for the various detector inefficiencies, the experimental cross section is extracted using Eq.~\ref{eq:xsecextract}. We then subtract the contribution to the cross section from $F_m(Q^2)$, using the parameterization of Ref.~\cite{AMROUN1994596} and applying an estimated 5\% systematic uncertainty on this value. This isolates the charge scattering contribution from which we extract the charge form factor, $F_c(Q^2)$. The $^3$H and $^3$He cross sections and charge form factors are given in Table~\ref{tab:results}. To extract the $^3$H/$^3$He charge form factor ratio, we take the ratio of the charge cross section, $\sigma_c^{dat}$, apply the statistical uncertainties along with the systematic uncertainties associated with the ratio from Tab.~\ref{tab:systematics}, yielding $F_c(^3\mathrm{H})/F_c(^3\mathrm{He}) = 1.2721 \pm 0.0132 \pm 0.0104$.

\begin{table}[tbh]
\begin{center}
\begin{tabular}{| c | l | l |}
\hline
                    &  $^3$H                            & $^3$He                           \\ \hline
$\sigma^{dat}$      &  $0.6264 \pm 0.0256 \pm 0.0063$   &   $1.2743 \pm 0.0533 \pm 0.0127$ \\
$\sigma_m^{fit}$    &  $0.1317 \pm 0.0066$              &   $0.0517 \pm 0.0026$            \\
$\sigma_c^{dat}$    &  $0.4946 \pm 0.0264 \pm 0.0063$   &   $1.2227 \pm 0.0534 \pm 0.0127$ \\
~$F_c^{dat}$~       &  $0.2555 \pm 0.0029 \pm 0.0021$   &   $0.2014 \pm 0.0044 \pm 0.0010$ \\ \hline
\end{tabular}
\caption{The cross section and charge form factor values extracted from
the measurements at $Q^2=3.03$~fm$^{-2}$. Cross sections are given in units of $\mu$b/sr. The first uncertainty is systematic and the second statistical, except for $\sigma_m^{fit}$, which is taken from a global fit. The corresponding ratios are $\sigma_c(^3\mathrm{H})/\sigma_c(^3\mathrm{He}) = 0.4046 \pm 0.0084 \pm 0.0066$ and $F_c(^3\mathrm{H})/F_c(^3\mathrm{He}) = 1.2721 \pm 0.0132 \pm 0.0104$, where the correlated systematic uncertainties largely cancel.}
\label{tab:results}
\end{center}
\end{table}

\section{Tritium Form Factor fit}

The linear falloff of $F_c(Q^2)$ associated with the RMS charge radius, as seen in Eq.~\ref{eq:radius}, is the dominant contribution from the nuclear structure up to roughly $Q^2=1$~fm$^{-2}$. Our precise measurement of the $^3$H/$^3$He $F_c$ ratio is at $Q^2 \approx 3$~fm$^{-2}$, and so has minimal direct impact on the extraction of the radius. However, because previous measurements often had large normalization uncertainties, even a single precise measurement can be used to help reduce the residual normalization uncertainty of the existing data.  We illustrate this by comparing fits to low-$Q^2$ measurements of the $^3$H charge form factor, with and without our new data, to determine the impact of the data. We do this by taking measurements of the $^3$H form factor that quoted values for $F_c(Q^2)$~\cite{collard65, beck82, Beck:1984qhv, Beck:1987zz}, and performing a global fit to these measurements, both with and without the inclusion of our new point. We then demonstrate the impact of lower-$Q^2$ measurements included in the E12-14-009 proposal~\cite{e1214009}.

\begin{figure}[htb]
\includegraphics[width=0.96\columnwidth, trim={10 1 0 5}, clip]{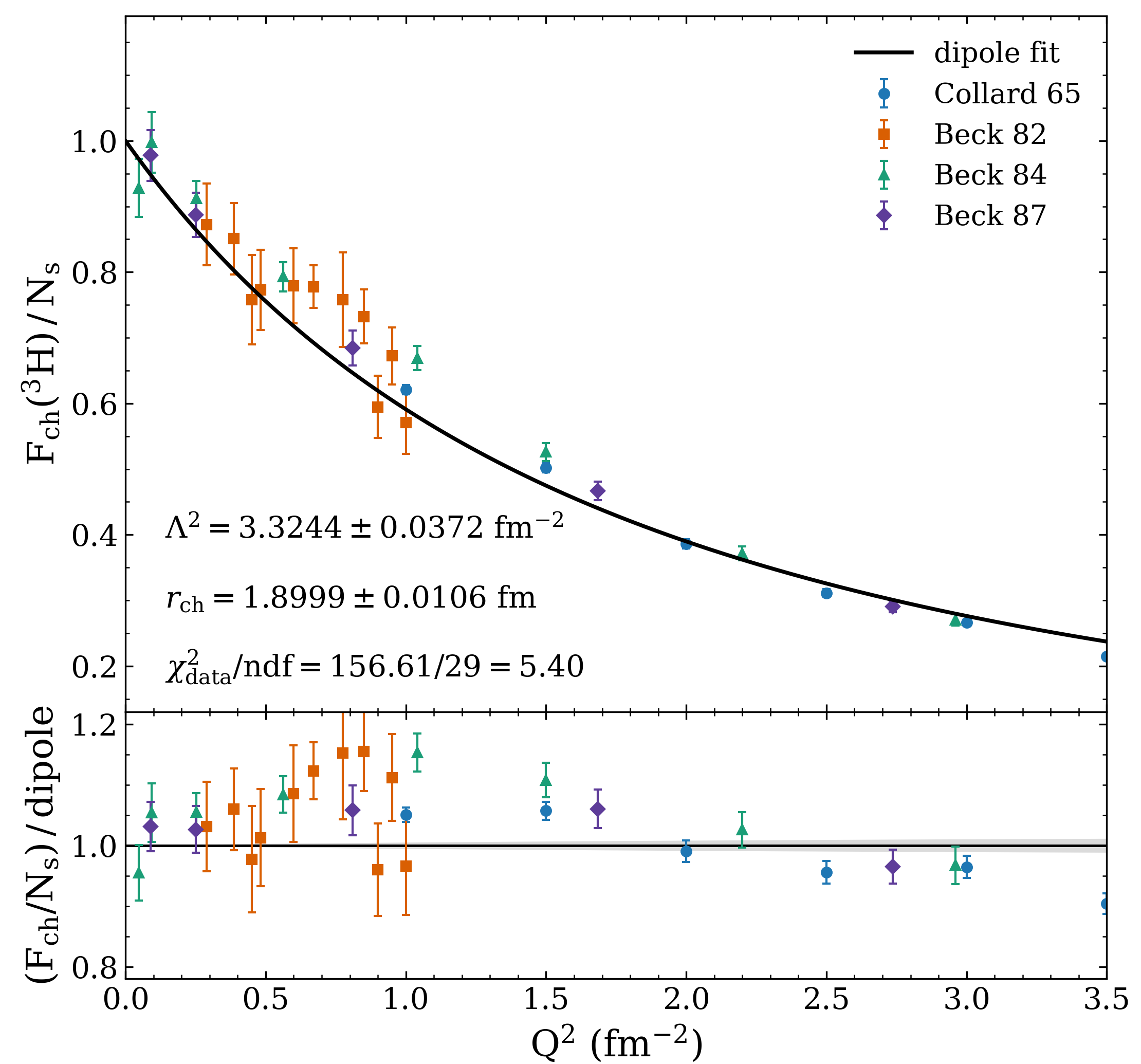}
 \caption{Top: Global fit to previous extractions of $F_c$~\cite{collard65, beck82, Beck:1984qhv, Beck:1987zz} using a dipole form factor, $F_{\rm ch}(Q^2) = (1+Q^2/\Lambda^2)^{-2}$, with the published uncertainties and a 1\% normalization uncertainty assigned to each data set. Data are plotted after dividing by the fitted normalization factor $N_s$. Bottom: Ratio of the data to the fit; the gray band indicates the uncertainty on the fit.}
 \label{fig:rawdata}
\end{figure}

Figure~\ref{fig:rawdata} shows the tritium $F_c(Q^2)$ results taken from Refs.~\cite{collard65, beck82, Beck:1984qhv, Beck:1987zz} compared to a simple dipole fit, along with the ratio of the data to the fit. Some data sets lie systematically above or below the fit, implying normalization offsets between the measurements. Although each experiment evaluated its systematic uncertainties, none separated normalization uncertainties from point-to-point (p2p) uncertainties, even for effects that would clearly impact all measurements the same way, e.g., the overall target thickness. We therefore examine each measurement and estimate which sources of uncertainty should be treated as normalization uncertainties. Some sources have both a common and an uncorrelated component; for example, the calibration of the beam current measurement affects all points equally, while time-dependent drift or variation of current between runs will have an effect that can differ for different measurements. When examining the sources of uncertainty, we estimate how much of the quoted uncertainty is common to all $Q^2$ values and how much varies from point to point. Finally, we add or increase sources of uncertainty where our modern understanding of effects such as radiative corrections warrants it.

For all four experiments, the normalization uncertainty is dominated by the determination of the amount of $^3$H in the target. Collard \etal~\cite{collard65} used high-pressure gas cells and measured the $^3$H cross sections relative to elastic scattering from hydrogen in an identical cell, so the absolute normalization depends on the gas densities determined from the equation of state and on the proton cross sections used as a reference. Beck \etal~\cite{Beck:1987zz} used cryogenic gas cells, for which the density determination was the largest source of uncertainty. The other two measurements used titanium tritide foil targets, for which the $^3$H areal density could not be measured directly. For Beck84~\cite{Beck:1984qhv}, the foil was wrinkled, and its absolute thickness was determined in a separate measurement of elastic proton scattering, giving a 4.1\% uncertainty that dominated that experiment. The normalization uncertainty is largest for Beck82~\cite{beck82}, where wrinkling of the foil during fabrication, combined with the small beam spot, made the number of target nuclei uncertain at the 7\% level; the target thickness was fixed by requiring the fitted form factor to extrapolate to $F_c(0) = 1$, and the resulting uncertainty was included in the quoted p2p uncertainties. Moving such contributions from the p2p to the normalization uncertainties reduces the p2p uncertainties, while the additional contributions described above increase them, yielding updated p2p uncertainties that are generally comparable to or somewhat larger than the published values. The data and uncertainties used in the fits are tabulated in the Supplemental Material~\cite{supplemental}, and these updated uncertainties are used in all subsequent fits.

For our sensitivity study, we combine our $^3$H/$^3$He charge form factor ratio measurement with the Amroun~\cite{AMROUN1994596} fit to the more-precisely-known $^3$He form factor to obtain a high-precision measurement of the $^3$H form factor at $Q^2\approx 3$~fm$^{-2}$. This point allows for better constraints on the normalization factor for the Collard~\cite{collard65} experiment, which covers the $Q^2$ range of this measurement. This in turn helps constrain normalization for the lower $Q^2$ measurements, where the data are more sensitive to the radius and the $\langle r^4 \rangle$ correction. To evaluate the potential impact of the new data, we perform a simple chi-squared fit to the data, allowing the normalization of each data set to float based on its estimated normalization uncertainties. Note that this is a simple sensitivity study and, as such, does not represent the full world data set, as we exclude experiments that quoted only cross sections or fits to the data rather than actual values of $F_c(Q^2)$. We also do not try to use more sophisticated fit functions that are generally considered more effective for a reliable extraction of the charge radius~\cite{AMROUN1994596, Sick:2003gm, Sick:2008zza, Hill:2010yb, Lee:2015jqa}. As such, this should not be considered an attempt to extract the $^3$H charge radius, rather it is a clean and simple test of the sensitivity of the radius extraction to a single precise measurement based on the $^3$H/$^3$He cross section ratio measurement presented here.

\begin{figure}[htb]
\includegraphics[width=0.96\columnwidth, trim={10 1 0 5}, clip]{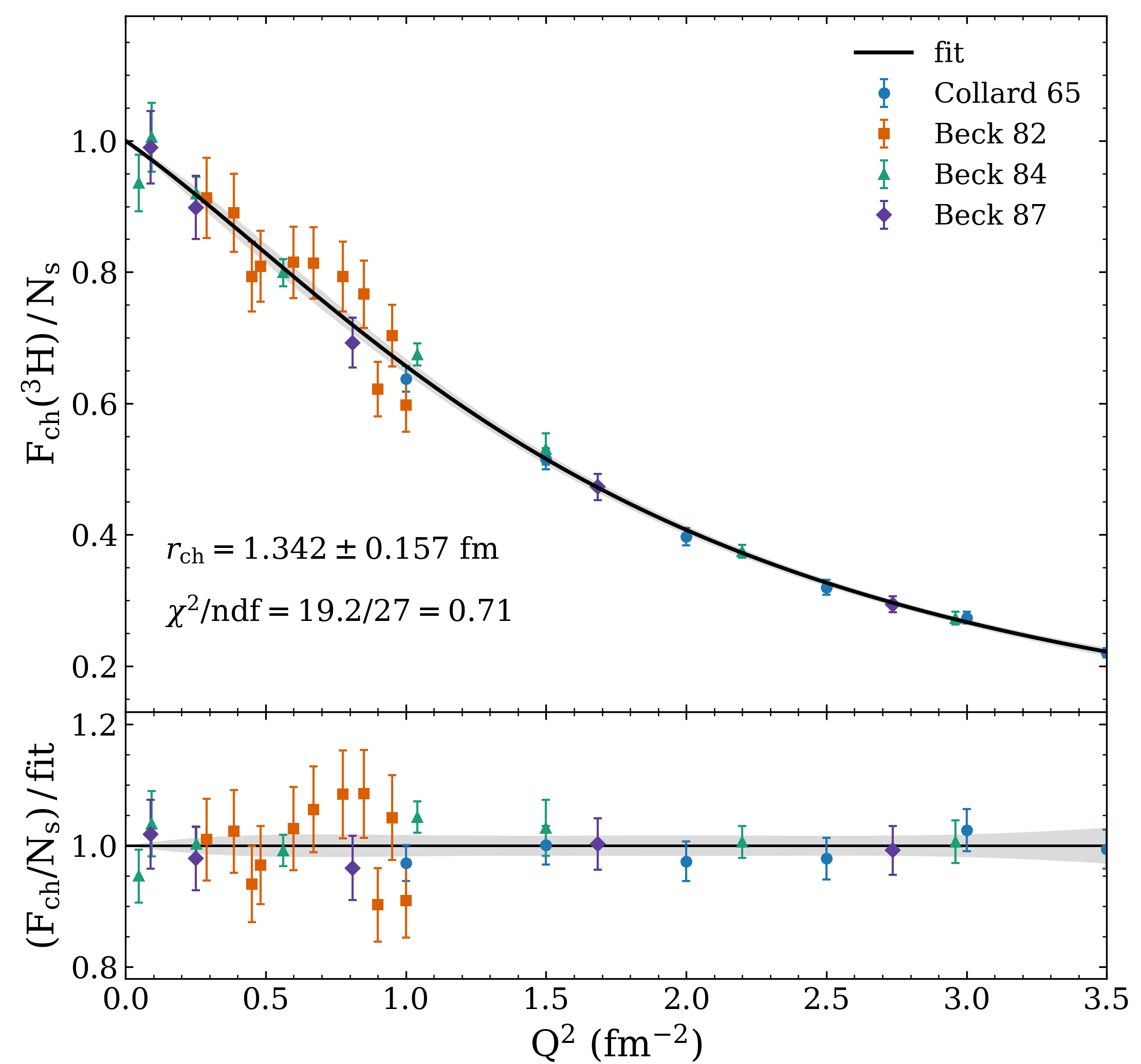} 
 \caption{Top: Global fit to previous extractions of $F_c$~\cite{collard65, beck82, Beck:1984qhv, Beck:1987zz} using an inverse 3rd order polynomial. When the data are plotted, the fitted normalization shift determined in the fit is applied to each data set. Bottom: Ratio of the data to the fit; the gray band indicates the uncertainty on the fit.}
 \label{fig:fit1}
\end{figure}

\begin{figure}[htb]
\includegraphics[width=0.96\columnwidth, trim={10 1 0 5}, clip]{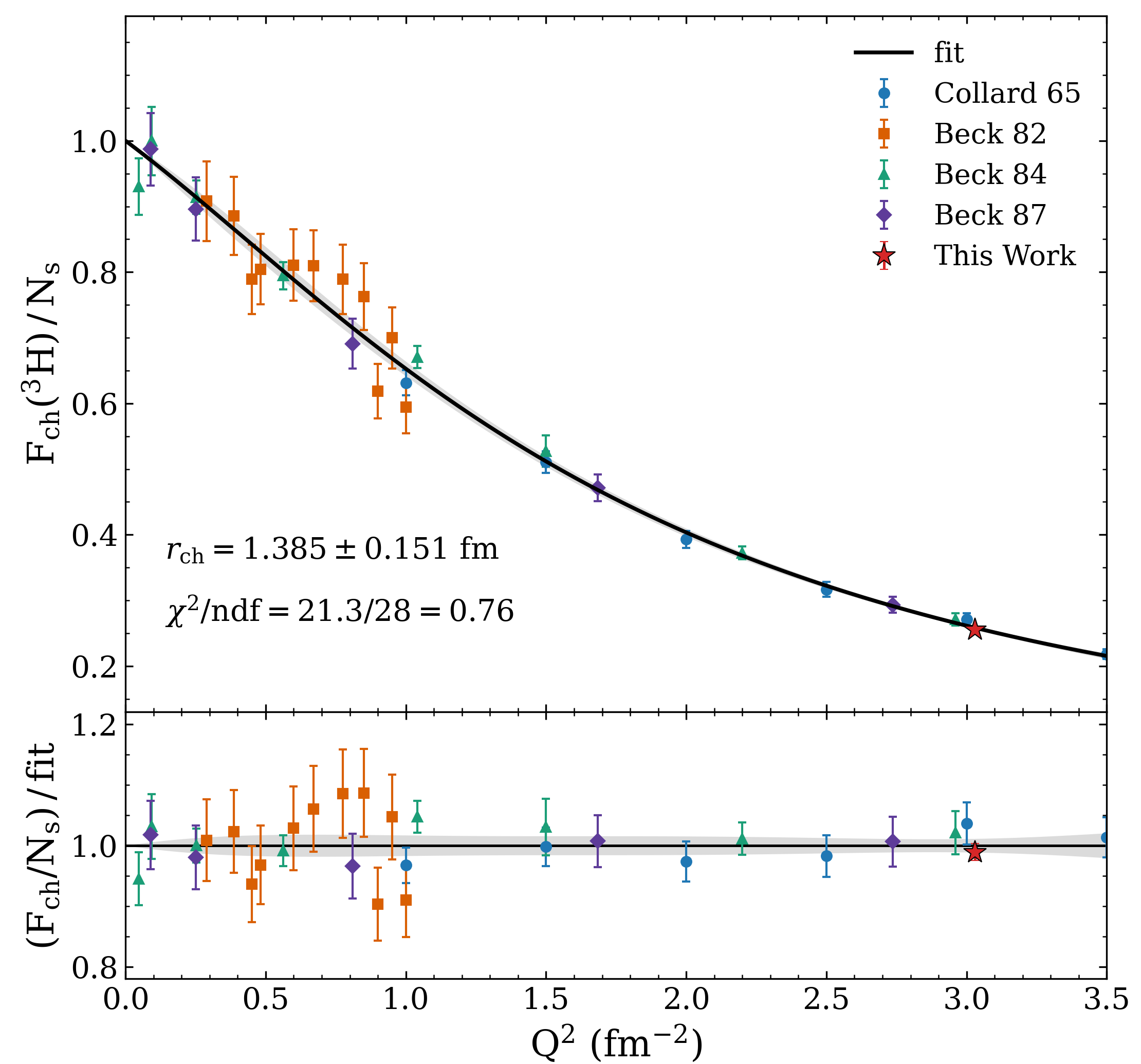} 
 \caption{Same as Fig.~\ref{fig:fit1}, but with the inclusion of our new high-precision point.}
 \label{fig:fit2}
\end{figure}

Figure~\ref{fig:fit1} shows the fit to the selected world data, using the updated systematic uncertainties as described above. The data are plotted after applying the normalization factor determined in the fit, and the bottom part of the plot shows the measurements divided by the final fit. Figure~\ref{fig:fit2} shows the fit when our new data point is included. Including our new data point has the greatest impact on the final normalization uncertainty of the Collard data, as this data set overlaps the $Q^2$ value of our point. Its uncertainty is reduced by 12\%, which provides smaller reductions ($\ltorder 5$\%) for the normalization factors of the other experiments. The radius uncertainty is only reduced by 4\%, as it is largely determined by the lower $Q^2$ data where the normalization uncertainties are not significantly reduced.  

\subsection{Projections for future measurements}

\begin{figure}[htb]
\includegraphics[width=0.96\columnwidth, trim={10 1 0 5}, clip]{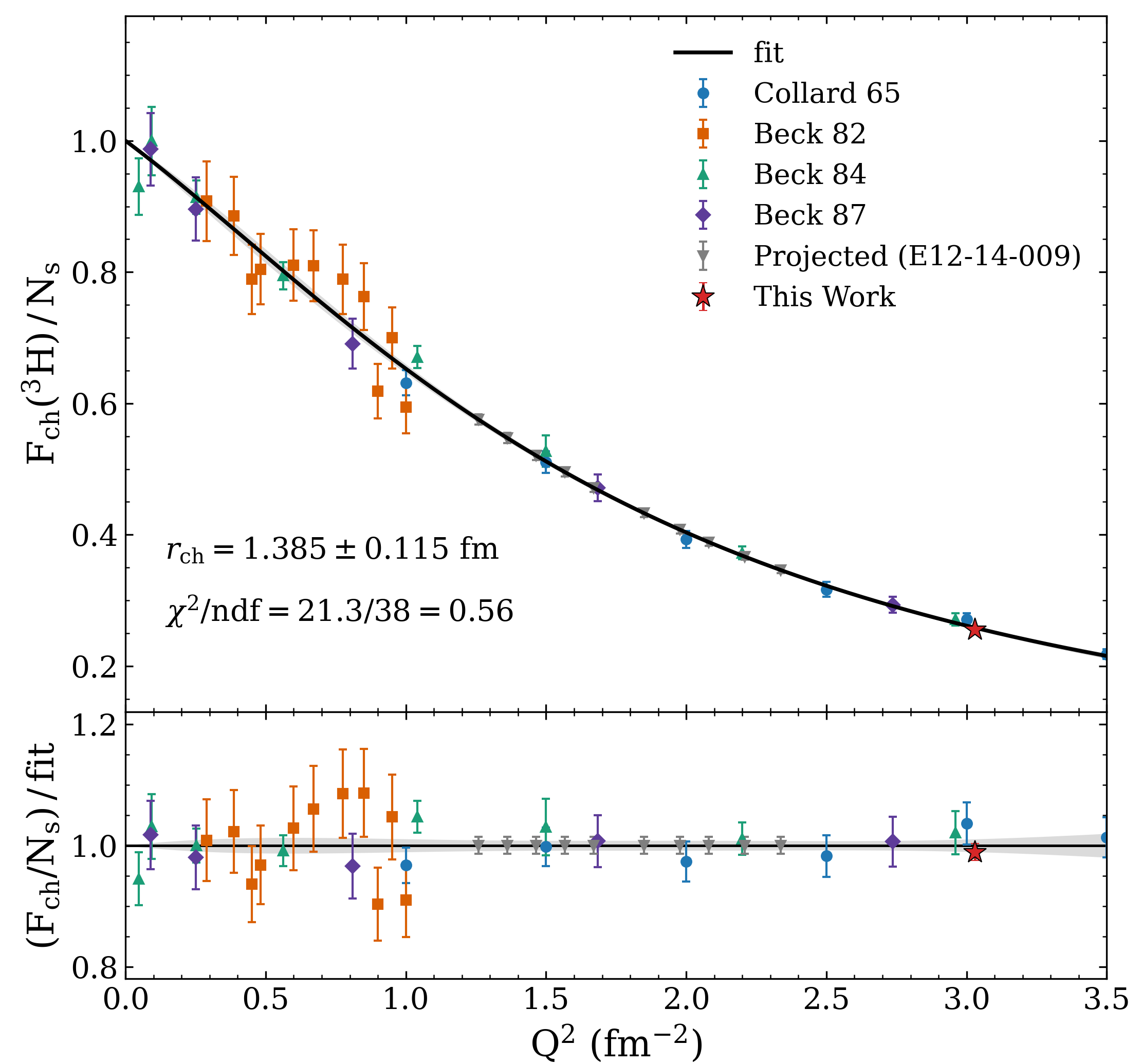} 
 \caption{Same as Fig.~\ref{fig:fit2}, but with the inclusion of the projected data for experiment E12-14-009~\cite{e1214009}. The uncertainties used in the fit are based on the uncertainties from the present data: a 1.4\% p2p uncertainty for each value of $F_c(Q^2)$ and a 0.78\% normalization uncertainty.}
 \label{fig:fit3}
\end{figure}

The proposal that motivated this measurement planned to measure several $Q^2$ values between 1.2 and 2.3~fm$^{-2}$, with a cross section ratio uncertainty of roughly 2\%~\cite{e1214009}. In this $Q^2$ range, the form factor is largely sensitive to the $\langle r^2 \rangle$ and $\langle r^4 \rangle$ terms, allowing for a global fit of all $^3$H and $^3$He data sets: cross sections, form factors, and target ratios. In the current measurement, only a single data point was obtained at a larger $Q^2$ ($Q^2 \approx 3$~fm$^{-2}$) than desired. As such, we perform a simple sensitivity study to estimate the impact of this technique on the extraction of the $^3$H charge radius, which limits the $^3$H-$^3$He charge radius difference. Figure~\ref{fig:fit3} shows the result of repeating the fit after including projected pseudo data based on the full proposed measurement. The pseudo data are placed on the fit from Fig.~\ref{fig:fit2}, so that they do not impact the total $\chi^2$ or the central value of the fit, but we can evaluate their impact on the uncertainties. 

Adding both our new point and the projected E12-14-009 data has a significantly larger impact on the radius. The radius uncertainty decreases from 0.157~fm to 0.115~fm, a 27\% reduction relative to the fit to the world data alone. The normalization uncertainties of the previous data sets are reduced by 10-20\% for Collard, Beck82, and Beck84, while Beck87 (which has the smallest normalization uncertainty and relatively large p2p uncertainties) is essentially unchanged. The projected points lie in the region $1.2 < Q^2 < 2.3$~fm$^{-2}$, where the form factor is most sensitive to $\langle r^2\rangle$ and $\langle r^4\rangle$, so they constrain the radius directly, in addition to the impact of improving the normalization of other data sets. This demonstrates the much greater value of high-precision ratio measurements at lower $Q^2$, as originally proposed.

We note that these numbers depend on the fit function used. When fitting the data to a 3rd order polynomial, adding our new data points reduces the improvement in both normalization and radius uncertainties by $\sim$50\%. When the projected data are added, the improvement in the normalization factor uncertainties is similar to the primary fit, but the uncertainty in the radius is reduced by almost a factor of two. In all cases, the inclusion of the proposed measurements at lower $Q^2$ has a significant impact on the extraction of the charge radius.

\begin{figure}[htb]
\includegraphics[width=0.96\columnwidth, trim={10 1 0 5}, clip]{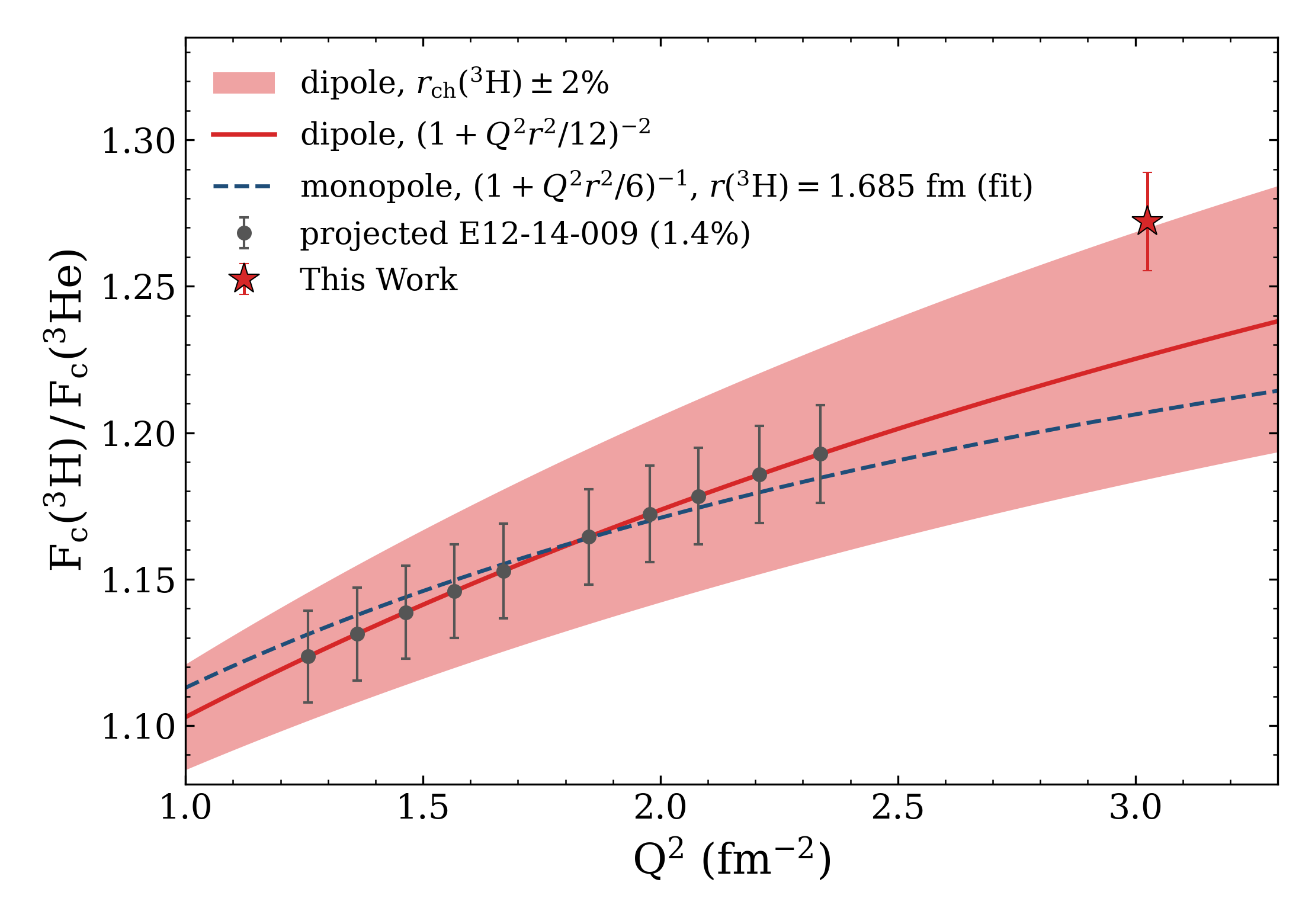}
 \caption{Ratio of the $^3$H and $^3$He charge form factors as a function of $Q^2$. The solid curve shows the ratio for dipole form factors, $(1+Q^2r^2/12)^{-2}$, calculated with the charge radii of Ref.~\cite{AMROUN1994596}, and the band shows the change in the dipole ratio for a $\pm2\%$ variation of the $^3$H radius. The dashed curve shows the ratio for monopole form factors, $(1+Q^2r^2/6)^{-1}$, with $r(^3\mathrm{He})$ fixed and $r(^3\mathrm{H}) = 1.685 \pm 0.008$~fm fitted to the projected points. The circles show the projected E12-14-009 measurements~\cite{e1214009}, placed on the dipole curve with 1.4\% p2p uncertainties (the 0.8\% normalization uncertainty is not shown). The star shows the present measurement at $Q^2 = 3.03$~fm$^{-2}$.}
\label{fig:dip_vs_mon}
\end{figure}

The sensitivity study above assumes that we combine the $^3$H/$^3$He form factor ratio with a global fit of the $^3$He form factor to extract an improved measurement for $^3$H, but there is another way to use such measurements to study the charge radii. The E12-14-009 proposal~\cite{e1214009} focused on using the form factor ratios to directly constrain the difference between the $^3$H and $^3$He charge radii, as seen in Eq.~\ref{eq:ratio}. This avoids relying on the previous form factor data and the $^3$He fits. As with the fits above, the higher-order terms in the $F_c(Q^2)$ expansion (eq.~\ref{eq:ffexpansion}) will yield corrections, but if the shape of the charge distributions is similar for $^3$H and $^3$He, the extraction of the charge radius difference is less sensitive to these higher-order terms. Figure~\ref{fig:dip_vs_mon} illustrates the sensitivity of this procedure to the details of the functional form of $F_c(Q^2)$.  The red curve is the value of the $^3$H/$^3$He ratio of $F_c(Q^2)$ assuming that both nuclei have a dipole form,  taking the $^3$H and $^3$He charge radii from Ref.~\cite{AMROUN1994596}, $r(^3\mathrm{H}) = 1.755$~fm and $r(^3\mathrm{He}) = 1.959$~fm, and the red band corresponds to a $\pm 2$\% change in the $^3$H radius. The blue dotted line corresponds to the same ratio if we assume a monopole form factor with the same radii. The central curves correspond to the form factor ratio for $\Delta R = 0.204$~fm, where $\Delta R$ is the difference between the $^3$H and $^3$He RMS charge radii, and the red shaded band corresponds to a change in $\Delta R$ of 0.035~fm. The data points for E12-14-009 are shown on the dipole curve with 1.4\% p2p uncertainties; the 0.8\% normalization uncertainty is not shown. In the region of the proposed measurements, the individual form factors differ by up to 20\% between the dipole and monopole forms, while the $^3$H/$^3$He ratios differ by less than 1\%.

If we were to assume that the charge distributions for $^3$H and $^3$He were the same except for the radius, i.e., the distribution is simply stretched out but not distorted going from $^3$H to $^3$He, the current data point would correspond to $\Delta R=0.293$~fm, with an uncertainty of roughly 0.003~fm and a similar uncertainty from the form factor dependence if we take the difference between the monopole and dipole form factor results at the same $Q^2$ value. However, at this $Q^2$ value the model dependence is large, and with only one data point we have no way to determine from the data to what extent the charge distributions are identical up to a rescaling in size. The projected E12-14-009 data provides a stronger constraint on $\Delta R$ from the experimental uncertainties, is less sensitive to the shape of the form factor under the assumption of rescaled charge distributions, and provides data over a range in $Q^2$ significant enough to test the assumption of identical charge distributions, given that this would introduce much larger deviations from the dipole or monopole predictions which agree well despite corresponding to very different charge distributions.

\section{Conclusions}

We have made a precise measurement of the $^3$H/$^3$He elastic cross section ratio and used this to extract the charge form factor ratio with a 1.32\% uncertainty. Combining this with a global fit to the $^3$He form factor provides a value for the $^3$H cross section with significantly smaller uncertainties than previous measurements. In a global fit, this improves the constraints on the relative experimental normalization factors but has minimal impact on the extraction of the $^3$H charge radius, as our new point is at relatively large $Q^2$ values. 

Future JLab measurements can significantly improve the measurement of the $^3$H-$^3$He charge radius difference. The current data only provides a significant improvement of this difference under the assumption that the two nuclei have charge distributions that are identical in shape and only differ in size. The new data will provide a sensitive test of this assumption and, assuming that deviations from this are small, will allow for a dramatically improved uncertainty in the difference between the $^3$H and $^3$He charge radii. In addition to being directly comparable to \textit{ab inito} structure calculations, it will provide a link between the precise atomic physics measurements of the radius difference between hydrogen isosopes and those between helium isotopes. 

\section{acknowledgments}

\begin{acknowledgments}

We acknowledge the contributions of the JLab target group and technical staff for the design and construction of the Tritium target and their support for this experiment. This work was supported in part by the Department of Energy's Office of Science, Office of Nuclear Physics, under contracts DE-AC02-05CH11231, DE-FG02-88ER40410, DE-SC0014168, DE-FG02-96ER40950, DE-SC0013615, and DE-SC0014168, the National Science Foundation including grant NSF PHY-1714809, the Science Committee of the Republic of Armenia under grant 21AG-1C085, the National Natural Science Foundation of China, under grant 12275148, the Tsinghua University Initiative Scientific Research Program, and DOE contract 89243126CSC000213 under which SURATech operates JLab. 

\end{acknowledgments}

\bibliographystyle{apsrev4-1}
\bibliography{tritium-NE}

\begin{thebibliography}{36}%
\makeatletter
\providecommand \@ifxundefined [1]{%
 \@ifx{#1\undefined}
}%
\providecommand \@ifnum [1]{%
 \ifnum #1\expandafter \@firstoftwo
 \else \expandafter \@secondoftwo
 \fi
}%
\providecommand \@ifx [1]{%
 \ifx #1\expandafter \@firstoftwo
 \else \expandafter \@secondoftwo
 \fi
}%
\providecommand \natexlab [1]{#1}%
\providecommand \enquote  [1]{``#1''}%
\providecommand \bibnamefont  [1]{#1}%
\providecommand \bibfnamefont [1]{#1}%
\providecommand \citenamefont [1]{#1}%
\providecommand \href@noop [0]{\@secondoftwo}%
\providecommand \href [0]{\begingroup \@sanitize@url \@href}%
\providecommand \@href[1]{\@@startlink{#1}\@@href}%
\providecommand \@@href[1]{\endgroup#1\@@endlink}%
\providecommand \@sanitize@url [0]{\catcode `\\12\catcode `\$12\catcode `\&12\catcode `\#12\catcode `\^12\catcode `\_12\catcode `\%12\relax}%
\providecommand \@@startlink[1]{}%
\providecommand \@@endlink[0]{}%
\providecommand \url  [0]{\begingroup\@sanitize@url \@url }%
\providecommand \@url [1]{\endgroup\@href {#1}{\urlprefix }}%
\providecommand \urlprefix  [0]{URL }%
\providecommand \Eprint [0]{\href }%
\providecommand \doibase [0]{http://dx.doi.org/}%
\providecommand \selectlanguage [0]{\@gobble}%
\providecommand \bibinfo  [0]{\@secondoftwo}%
\providecommand \bibfield  [0]{\@secondoftwo}%
\providecommand \translation [1]{[#1]}%
\providecommand \BibitemOpen [0]{}%
\providecommand \bibitemStop [0]{}%
\providecommand \bibitemNoStop [0]{.\EOS\space}%
\providecommand \EOS [0]{\spacefactor3000\relax}%
\providecommand \BibitemShut  [1]{\csname bibitem#1\endcsname}%
\let\auto@bib@innerbib\@empty
\bibitem [{\citenamefont {Rosenbluth}(1950)}]{Rosenbluth:1950yq}%
  \BibitemOpen
  \bibfield  {author} {\bibinfo {author} {\bibfnamefont {M.~N.}\ \bibnamefont {Rosenbluth}},\ }\href {\doibase 10.1103/PhysRev.79.615} {\bibfield  {journal} {\bibinfo  {journal} {Phys. Rev.}\ }\textbf {\bibinfo {volume} {79}},\ \bibinfo {pages} {615} (\bibinfo {year} {1950})}\BibitemShut {NoStop}%
\bibitem [{\citenamefont {Miller}(2009)}]{Miller:2009sg}%
  \BibitemOpen
  \bibfield  {author} {\bibinfo {author} {\bibfnamefont {G.~A.}\ \bibnamefont {Miller}},\ }\href {\doibase 10.1103/PhysRevC.80.045210} {\bibfield  {journal} {\bibinfo  {journal} {Phys. Rev. C}\ }\textbf {\bibinfo {volume} {80}},\ \bibinfo {pages} {045210} (\bibinfo {year} {2009})}\BibitemShut {NoStop}%
\bibitem [{\citenamefont {Jentschura}\ \emph {et~al.}(2011)\citenamefont {Jentschura}, \citenamefont {Matveev}, \citenamefont {Parthey}, \citenamefont {Alnis}, \citenamefont {Pohl}, \citenamefont {Udem}, \citenamefont {Kolachevsky},\ and\ \citenamefont {H{\"a}nsch}}]{Jentschura:2011mdl}%
  \BibitemOpen
  \bibfield  {author} {\bibinfo {author} {\bibfnamefont {U.~D.}\ \bibnamefont {Jentschura}}, \bibinfo {author} {\bibfnamefont {A.}~\bibnamefont {Matveev}}, \bibinfo {author} {\bibfnamefont {C.~G.}\ \bibnamefont {Parthey}}, \bibinfo {author} {\bibfnamefont {J.}~\bibnamefont {Alnis}}, \bibinfo {author} {\bibfnamefont {R.}~\bibnamefont {Pohl}}, \bibinfo {author} {\bibfnamefont {T.}~\bibnamefont {Udem}}, \bibinfo {author} {\bibfnamefont {N.}~\bibnamefont {Kolachevsky}}, \ and\ \bibinfo {author} {\bibfnamefont {T.~W.}\ \bibnamefont {H{\"a}nsch}},\ }\href {\doibase 10.1103/PhysRevA.83.042505} {\bibfield  {journal} {\bibinfo  {journal} {Phys. Rev. A}\ }\textbf {\bibinfo {volume} {83}},\ \bibinfo {pages} {042505} (\bibinfo {year} {2011})}\BibitemShut {NoStop}%
\bibitem [{\citenamefont {Mueller}\ \emph {et~al.}(2007)\citenamefont {Mueller} \emph {et~al.}}]{Mueller:2007dhq}%
  \BibitemOpen
  \bibfield  {author} {\bibinfo {author} {\bibfnamefont {P.}~\bibnamefont {Mueller}} \emph {et~al.},\ }\href {\doibase 10.1103/PhysRevLett.99.252501} {\bibfield  {journal} {\bibinfo  {journal} {Phys. Rev. Lett.}\ }\textbf {\bibinfo {volume} {99}},\ \bibinfo {pages} {252501} (\bibinfo {year} {2007})}\BibitemShut {NoStop}%
\bibitem [{\citenamefont {Mohr}\ \emph {et~al.}(2025)\citenamefont {Mohr}, \citenamefont {Newell}, \citenamefont {Taylor},\ and\ \citenamefont {Tiesinga}}]{Mohr:2024kco}%
  \BibitemOpen
  \bibfield  {author} {\bibinfo {author} {\bibfnamefont {P.~J.}\ \bibnamefont {Mohr}}, \bibinfo {author} {\bibfnamefont {D.~B.}\ \bibnamefont {Newell}}, \bibinfo {author} {\bibfnamefont {B.~N.}\ \bibnamefont {Taylor}}, \ and\ \bibinfo {author} {\bibfnamefont {E.}~\bibnamefont {Tiesinga}},\ }\href {\doibase 10.1103/RevModPhys.97.025002} {\bibfield  {journal} {\bibinfo  {journal} {Rev. Mod. Phys.}\ }\textbf {\bibinfo {volume} {97}},\ \bibinfo {pages} {025002} (\bibinfo {year} {2025})}\BibitemShut {NoStop}%
\bibitem [{\citenamefont {Collard}\ \emph {et~al.}(1965)\citenamefont {Collard}, \citenamefont {Hofstadter}, \citenamefont {Hughes}, \citenamefont {Johansson}, \citenamefont {Yearian}, \citenamefont {Day},\ and\ \citenamefont {Wagner}}]{collard65}%
  \BibitemOpen
  \bibfield  {author} {\bibinfo {author} {\bibfnamefont {H.}~\bibnamefont {Collard}}, \bibinfo {author} {\bibfnamefont {R.}~\bibnamefont {Hofstadter}}, \bibinfo {author} {\bibfnamefont {E.~B.}\ \bibnamefont {Hughes}}, \bibinfo {author} {\bibfnamefont {A.}~\bibnamefont {Johansson}}, \bibinfo {author} {\bibfnamefont {M.~R.}\ \bibnamefont {Yearian}}, \bibinfo {author} {\bibfnamefont {R.~B.}\ \bibnamefont {Day}}, \ and\ \bibinfo {author} {\bibfnamefont {R.~T.}\ \bibnamefont {Wagner}},\ }\href@noop {} {\bibfield  {journal} {\bibinfo  {journal} {Phys. Rev.}\ }\textbf {\bibinfo {volume} {138}},\ \bibinfo {pages} {B57} (\bibinfo {year} {1965})}\BibitemShut {NoStop}%
\bibitem [{\citenamefont {Beck}\ \emph {et~al.}(1989)\citenamefont {Beck} \emph {et~al.}}]{Beck:1989bi}%
  \BibitemOpen
  \bibfield  {author} {\bibinfo {author} {\bibfnamefont {D.}~\bibnamefont {Beck}} \emph {et~al.},\ }\href@noop {} {\bibfield  {journal} {\bibinfo  {journal} {Nucl. Instrum. Meth. A}\ }\textbf {\bibinfo {volume} {277}},\ \bibinfo {pages} {323} (\bibinfo {year} {1989})}\BibitemShut {NoStop}%
\bibitem [{\citenamefont {Amroun}\ \emph {et~al.}(1994)\citenamefont {Amroun} \emph {et~al.}}]{AMROUN1994596}%
  \BibitemOpen
  \bibfield  {author} {\bibinfo {author} {\bibfnamefont {A.}~\bibnamefont {Amroun}} \emph {et~al.},\ }\href {\doibase 10.1016/0375-9474(94)90925-3} {\bibfield  {journal} {\bibinfo  {journal} {Nucl. Phys. A}\ }\textbf {\bibinfo {volume} {579}},\ \bibinfo {pages} {596} (\bibinfo {year} {1994})}\BibitemShut {NoStop}%
\bibitem [{\citenamefont {Arrington}\ \emph {et~al.}(2011)\citenamefont {Arrington}, \citenamefont {Day}, \citenamefont {Higinbotham},\ and\ \citenamefont {Solvignon}}]{e1211112}%
  \BibitemOpen
  \bibfield  {author} {\bibinfo {author} {\bibfnamefont {J.}~\bibnamefont {Arrington}}, \bibinfo {author} {\bibfnamefont {D.}~\bibnamefont {Day}}, \bibinfo {author} {\bibfnamefont {D.~W.}\ \bibnamefont {Higinbotham}}, \ and\ \bibinfo {author} {\bibfnamefont {P.}~\bibnamefont {Solvignon}},\ }\href@noop {} {\enquote {\bibinfo {title} {{Precision measurement of the isospin dependence in the 2N and 3N short range correlation region}},}\ }\bibinfo {howpublished} {Jefferson Lab Experiment Proposal E12-11-112} (\bibinfo {year} {2011})\BibitemShut {NoStop}%
\bibitem [{\citenamefont {Arrington}\ \emph {et~al.}(2023)\citenamefont {Arrington}, \citenamefont {Cruz-Torres}, \citenamefont {Hague}, \citenamefont {Kurbany}, \citenamefont {Li}, \citenamefont {Meekins},\ and\ \citenamefont {Santiesteban}}]{Arrington:2023hht}%
  \BibitemOpen
  \bibfield  {author} {\bibinfo {author} {\bibfnamefont {J.}~\bibnamefont {Arrington}}, \bibinfo {author} {\bibfnamefont {R.}~\bibnamefont {Cruz-Torres}}, \bibinfo {author} {\bibfnamefont {T.~J.}\ \bibnamefont {Hague}}, \bibinfo {author} {\bibfnamefont {L.}~\bibnamefont {Kurbany}}, \bibinfo {author} {\bibfnamefont {S.}~\bibnamefont {Li}}, \bibinfo {author} {\bibfnamefont {D.}~\bibnamefont {Meekins}}, \ and\ \bibinfo {author} {\bibfnamefont {N.}~\bibnamefont {Santiesteban}},\ }\href {\doibase 10.1140/epja/s10050-023-01085-6} {\bibfield  {journal} {\bibinfo  {journal} {Eur. Phys. J. A}\ }\textbf {\bibinfo {volume} {59}},\ \bibinfo {pages} {188} (\bibinfo {year} {2023})}\BibitemShut {NoStop}%
\bibitem [{\citenamefont {Li}\ \emph {et~al.}(2022)\citenamefont {Li} \emph {et~al.}}]{Li:2022fhh}%
  \BibitemOpen
  \bibfield  {author} {\bibinfo {author} {\bibfnamefont {S.}~\bibnamefont {Li}} \emph {et~al.},\ }\href {\doibase 10.1038/s41586-022-05007-2} {\bibfield  {journal} {\bibinfo  {journal} {Nature}\ }\textbf {\bibinfo {volume} {609}},\ \bibinfo {pages} {41} (\bibinfo {year} {2022})}\BibitemShut {NoStop}%
\bibitem [{\citenamefont {Santiesteban}\ \emph {et~al.}(2024)\citenamefont {Santiesteban} \emph {et~al.}}]{Santiesteban:2023rsh}%
  \BibitemOpen
  \bibfield  {author} {\bibinfo {author} {\bibfnamefont {S.~N.}\ \bibnamefont {Santiesteban}} \emph {et~al.} (\bibinfo {collaboration} {Jefferson Lab Hall A}),\ }\href {\doibase 10.1103/PhysRevLett.132.162501} {\bibfield  {journal} {\bibinfo  {journal} {Phys. Rev. Lett.}\ }\textbf {\bibinfo {volume} {132}},\ \bibinfo {pages} {162501} (\bibinfo {year} {2024})}\BibitemShut {NoStop}%
\bibitem [{\citenamefont {Li}\ \emph {et~al.}(2025)\citenamefont {Li} \emph {et~al.}}]{Li:2024rzf}%
  \BibitemOpen
  \bibfield  {author} {\bibinfo {author} {\bibfnamefont {S.}~\bibnamefont {Li}} \emph {et~al.},\ }\href {\doibase 10.1016/j.physletb.2025.139734} {\bibfield  {journal} {\bibinfo  {journal} {Phys. Lett. B}\ }\textbf {\bibinfo {volume} {868}},\ \bibinfo {pages} {139734} (\bibinfo {year} {2025})}\BibitemShut {NoStop}%
\bibitem [{\citenamefont {Arrington}\ \emph {et~al.}(2014)\citenamefont {Arrington}, \citenamefont {Averett}, \citenamefont {Higinbotham},\ and\ \citenamefont {Myers}}]{e1214009}%
  \BibitemOpen
  \bibfield  {author} {\bibinfo {author} {\bibfnamefont {J.}~\bibnamefont {Arrington}}, \bibinfo {author} {\bibfnamefont {T.}~\bibnamefont {Averett}}, \bibinfo {author} {\bibfnamefont {D.~W.}\ \bibnamefont {Higinbotham}}, \ and\ \bibinfo {author} {\bibfnamefont {L.}~\bibnamefont {Myers}},\ }\href@noop {} {\enquote {\bibinfo {title} {{Ratio of the electric form factor in the mirror nuclei 3He and 3H}},}\ }\bibinfo {howpublished} {Jefferson Lab Experiment Proposal E12-14-009} (\bibinfo {year} {2014})\BibitemShut {NoStop}%
\bibitem [{\citenamefont {Leemann}\ \emph {et~al.}(2001)\citenamefont {Leemann}, \citenamefont {Douglas},\ and\ \citenamefont {Krafft}}]{Leemann:2001dg}%
  \BibitemOpen
  \bibfield  {author} {\bibinfo {author} {\bibfnamefont {C.~W.}\ \bibnamefont {Leemann}}, \bibinfo {author} {\bibfnamefont {D.~R.}\ \bibnamefont {Douglas}}, \ and\ \bibinfo {author} {\bibfnamefont {G.~A.}\ \bibnamefont {Krafft}},\ }\href {\doibase 10.1146/annurev.nucl.51.101701.132327} {\bibfield  {journal} {\bibinfo  {journal} {Ann. Rev. Nucl. Part. Sci.}\ }\textbf {\bibinfo {volume} {51}},\ \bibinfo {pages} {413} (\bibinfo {year} {2001})}\BibitemShut {NoStop}%
\bibitem [{\citenamefont {Alcorn}\ \emph {et~al.}(2004)\citenamefont {Alcorn} \emph {et~al.}}]{Alcorn:2004sb}%
  \BibitemOpen
  \bibfield  {author} {\bibinfo {author} {\bibfnamefont {J.}~\bibnamefont {Alcorn}} \emph {et~al.},\ }\href {\doibase 10.1016/j.nima.2003.11.415} {\bibfield  {journal} {\bibinfo  {journal} {Nucl. Instrum. Meth. A}\ }\textbf {\bibinfo {volume} {522}},\ \bibinfo {pages} {294} (\bibinfo {year} {2004})}\BibitemShut {NoStop}%
\bibitem [{\citenamefont {Brajuskovic}\ \emph {et~al.}(2013)\citenamefont {Brajuskovic}, \citenamefont {O'Connor}, \citenamefont {Holt}, \citenamefont {Reneker}, \citenamefont {Meekins},\ and\ \citenamefont {Solvignon}}]{target_NIM}%
  \BibitemOpen
  \bibfield  {author} {\bibinfo {author} {\bibfnamefont {B.}~\bibnamefont {Brajuskovic}}, \bibinfo {author} {\bibfnamefont {T.}~\bibnamefont {O'Connor}}, \bibinfo {author} {\bibfnamefont {R.}~\bibnamefont {Holt}}, \bibinfo {author} {\bibfnamefont {J.}~\bibnamefont {Reneker}}, \bibinfo {author} {\bibfnamefont {D.}~\bibnamefont {Meekins}}, \ and\ \bibinfo {author} {\bibfnamefont {P.}~\bibnamefont {Solvignon}},\ }\href {\doibase 10.1016/j.nima.2013.06.090} {\bibfield  {journal} {\bibinfo  {journal} {Nucl. Instrum. Meth. A}\ }\textbf {\bibinfo {volume} {729}},\ \bibinfo {pages} {469} (\bibinfo {year} {2013})}\BibitemShut {NoStop}%
\bibitem [{\citenamefont {Abrams}\ \emph {et~al.}(2022)\citenamefont {Abrams} \emph {et~al.}}]{Abrams:2021usd}%
  \BibitemOpen
  \bibfield  {author} {\bibinfo {author} {\bibfnamefont {D.}~\bibnamefont {Abrams}} \emph {et~al.} (\bibinfo {collaboration} {Jefferson Lab Hall A Tritium}),\ }\href {\doibase 10.1103/PhysRevLett.128.132003} {\bibfield  {journal} {\bibinfo  {journal} {Phys. Rev. Lett.}\ }\textbf {\bibinfo {volume} {128}},\ \bibinfo {pages} {132003} (\bibinfo {year} {2022})}\BibitemShut {NoStop}%
\bibitem [{\citenamefont {Santiesteban}\ \emph {et~al.}(2019)\citenamefont {Santiesteban} \emph {et~al.}}]{Santiesteban:2018qwi}%
  \BibitemOpen
  \bibfield  {author} {\bibinfo {author} {\bibfnamefont {S.~N.}\ \bibnamefont {Santiesteban}} \emph {et~al.},\ }\href {\doibase 10.1016/j.nima.2019.06.025} {\bibfield  {journal} {\bibinfo  {journal} {Nucl. Instrum. Meth. A}\ }\textbf {\bibinfo {volume} {940}},\ \bibinfo {pages} {351} (\bibinfo {year} {2019})}\BibitemShut {NoStop}%
\bibitem [{\citenamefont {Kurbani}(2023)}]{kurbani_thesis}%
  \BibitemOpen
  \bibfield  {author} {\bibinfo {author} {\bibfnamefont {L.}~\bibnamefont {Kurbani}},\ }\href@noop {} {\enquote {\bibinfo {title} {{Ph.D Thesis, University New Hampshire}},}\ } (\bibinfo {year} {2023})\BibitemShut {NoStop}%
\bibitem [{\citenamefont {Ou}(2019)}]{Ou:2019ing}%
  \BibitemOpen
  \bibfield  {author} {\bibinfo {author} {\bibfnamefont {L.}~\bibnamefont {Ou}},\ }\emph {\bibinfo {title} {{Precision Measurements of Electron-Proton Elastic Scattering Cross Sections at Large $Q^2$}}},\ \href {\doibase 10.2172/1595240} {Ph.D. thesis},\ \bibinfo  {school} {MIT, Cambridge, Dept. Phys.} (\bibinfo {year} {2019})\BibitemShut {NoStop}%
\bibitem [{\citenamefont {Cruz-Torres}\ \emph {et~al.}(2019)\citenamefont {Cruz-Torres} \emph {et~al.}}]{cruz-torres19}%
  \BibitemOpen
  \bibfield  {author} {\bibinfo {author} {\bibfnamefont {R.}~\bibnamefont {Cruz-Torres}} \emph {et~al.},\ }\href {\doibase 10.1016/j.physletb.2019.134890} {\bibfield  {journal} {\bibinfo  {journal} {Phys. Lett. B}\ }\textbf {\bibinfo {volume} {797}},\ \bibinfo {pages} {134890} (\bibinfo {year} {2019})}\BibitemShut {NoStop}%
\bibitem [{\citenamefont {Cruz-Torres}\ \emph {et~al.}(2020)\citenamefont {Cruz-Torres} \emph {et~al.}}]{cruz-torres20}%
  \BibitemOpen
  \bibfield  {author} {\bibinfo {author} {\bibfnamefont {R.}~\bibnamefont {Cruz-Torres}} \emph {et~al.},\ }\href {\doibase 10.1103/PhysRevLett.124.212501} {\bibfield  {journal} {\bibinfo  {journal} {Phys. Rev. Lett.}\ }\textbf {\bibinfo {volume} {124}},\ \bibinfo {pages} {212501} (\bibinfo {year} {2020})}\BibitemShut {NoStop}%
\bibitem [{\citenamefont {Abrams}\ \emph {et~al.}(2025)\citenamefont {Abrams} \emph {et~al.}}]{Abrams:2024las}%
  \BibitemOpen
  \bibfield  {author} {\bibinfo {author} {\bibfnamefont {D.}~\bibnamefont {Abrams}} \emph {et~al.} (\bibinfo {collaboration} {Jefferson Lab Hall A Tritium}),\ }\href {\doibase 10.1103/31xz-s84d} {\bibfield  {journal} {\bibinfo  {journal} {Phys. Rev. Lett.}\ }\textbf {\bibinfo {volume} {135}},\ \bibinfo {pages} {062502} (\bibinfo {year} {2025})}\BibitemShut {NoStop}%
\bibitem [{\citenamefont {Gaskell}(2001)}]{Gaskell:2001}%
  \BibitemOpen
  \bibfield  {author} {\bibinfo {author} {\bibfnamefont {D.~J.}\ \bibnamefont {Gaskell}},\ }\emph {\bibinfo {title} {{Longitudinal Electroproduction of Charged Pions from Hydrogen, Deuterium, and Helium-3}}},\ \href@noop {} {\bibinfo {type} {{Ph.D. Thesis}}},\ \bibinfo  {school} {Oregon State University} (\bibinfo {year} {2001})\BibitemShut {NoStop}%
\bibitem [{\citenamefont {Mohring}\ \emph {et~al.}(2003)\citenamefont {Mohring} \emph {et~al.}}]{Mohring:2002cpu}%
  \BibitemOpen
  \bibfield  {author} {\bibinfo {author} {\bibfnamefont {R.~M.}\ \bibnamefont {Mohring}} \emph {et~al.},\ }\href {\doibase 10.1103/PhysRevC.67.055205} {\bibfield  {journal} {\bibinfo  {journal} {Phys. Rev. C}\ }\textbf {\bibinfo {volume} {67}},\ \bibinfo {pages} {055205} (\bibinfo {year} {2003})}\BibitemShut {NoStop}%
\bibitem [{\citenamefont {Ent}\ \emph {et~al.}(2001)\citenamefont {Ent}, \citenamefont {Filippone}, \citenamefont {Makins}, \citenamefont {Milner}, \citenamefont {O'Neill},\ and\ \citenamefont {Wasson}}]{Ent:2001hm}%
  \BibitemOpen
  \bibfield  {author} {\bibinfo {author} {\bibfnamefont {R.}~\bibnamefont {Ent}}, \bibinfo {author} {\bibfnamefont {B.~W.}\ \bibnamefont {Filippone}}, \bibinfo {author} {\bibfnamefont {N.~C.~R.}\ \bibnamefont {Makins}}, \bibinfo {author} {\bibfnamefont {R.~G.}\ \bibnamefont {Milner}}, \bibinfo {author} {\bibfnamefont {T.~G.}\ \bibnamefont {O'Neill}}, \ and\ \bibinfo {author} {\bibfnamefont {D.~A.}\ \bibnamefont {Wasson}},\ }\href {\doibase 10.1103/PhysRevC.64.054610} {\bibfield  {journal} {\bibinfo  {journal} {Phys. Rev. C}\ }\textbf {\bibinfo {volume} {64}},\ \bibinfo {pages} {054610} (\bibinfo {year} {2001})}\BibitemShut {NoStop}%
\bibitem [{\citenamefont {Aste}\ \emph {et~al.}(2005)\citenamefont {Aste}, \citenamefont {von Arx},\ and\ \citenamefont {Trautmann}}]{Aste:2007hh}%
  \BibitemOpen
  \bibfield  {author} {\bibinfo {author} {\bibfnamefont {A.}~\bibnamefont {Aste}}, \bibinfo {author} {\bibfnamefont {C.}~\bibnamefont {von Arx}}, \ and\ \bibinfo {author} {\bibfnamefont {D.}~\bibnamefont {Trautmann}},\ }\href {\doibase 10.1140/epja/i2005-10159-x} {\bibfield  {journal} {\bibinfo  {journal} {Eur. Phys. J. A}\ }\textbf {\bibinfo {volume} {26}},\ \bibinfo {pages} {167} (\bibinfo {year} {2005})}\BibitemShut {NoStop}%
\bibitem [{\citenamefont {Beck}\ \emph {et~al.}(1982)\citenamefont {Beck}, \citenamefont {Asai},\ and\ \citenamefont {Skopik}}]{beck82}%
  \BibitemOpen
  \bibfield  {author} {\bibinfo {author} {\bibfnamefont {D.}~\bibnamefont {Beck}}, \bibinfo {author} {\bibfnamefont {J.}~\bibnamefont {Asai}}, \ and\ \bibinfo {author} {\bibfnamefont {D.}~\bibnamefont {Skopik}},\ }\href@noop {} {\bibfield  {journal} {\bibinfo  {journal} {Physical Review C}\ }\textbf {\bibinfo {volume} {25}},\ \bibinfo {pages} {1152} (\bibinfo {year} {1982})}\BibitemShut {NoStop}%
\bibitem [{\citenamefont {Beck}\ \emph {et~al.}(1984)\citenamefont {Beck} \emph {et~al.}}]{Beck:1984qhv}%
  \BibitemOpen
  \bibfield  {author} {\bibinfo {author} {\bibfnamefont {D.~H.}\ \bibnamefont {Beck}} \emph {et~al.},\ }\href@noop {} {\bibfield  {journal} {\bibinfo  {journal} {Phys. Rev. C}\ }\textbf {\bibinfo {volume} {30}},\ \bibinfo {pages} {1403} (\bibinfo {year} {1984})}\BibitemShut {NoStop}%
\bibitem [{\citenamefont {Beck}\ \emph {et~al.}(1987)\citenamefont {Beck} \emph {et~al.}}]{Beck:1987zz}%
  \BibitemOpen
  \bibfield  {author} {\bibinfo {author} {\bibfnamefont {D.}~\bibnamefont {Beck}} \emph {et~al.},\ }\href {\doibase 10.1103/PhysRevLett.59.1537} {\bibfield  {journal} {\bibinfo  {journal} {Phys. Rev. Lett.}\ }\textbf {\bibinfo {volume} {59}},\ \bibinfo {pages} {1537} (\bibinfo {year} {1987})}\BibitemShut {NoStop}%
\bibitem [{sup()}]{supplemental}%
  \BibitemOpen
  \href@noop {} {}\bibinfo {note} {See Supplemental Material at [URL will be inserted by publisher] for detailed expressions and tables of cross section results.}\BibitemShut {Stop}%
\bibitem [{\citenamefont {Sick}(2003)}]{Sick:2003gm}%
  \BibitemOpen
  \bibfield  {author} {\bibinfo {author} {\bibfnamefont {I.}~\bibnamefont {Sick}},\ }\href {\doibase 10.1016/j.physletb.2003.09.092} {\bibfield  {journal} {\bibinfo  {journal} {Phys. Lett. B}\ }\textbf {\bibinfo {volume} {576}},\ \bibinfo {pages} {62} (\bibinfo {year} {2003})}\BibitemShut {NoStop}%
\bibitem [{\citenamefont {Sick}(2008)}]{Sick:2008zza}%
  \BibitemOpen
  \bibfield  {author} {\bibinfo {author} {\bibfnamefont {I.}~\bibnamefont {Sick}},\ }\href {\doibase 10.1103/PhysRevC.77.041302} {\bibfield  {journal} {\bibinfo  {journal} {Phys. Rev. C}\ }\textbf {\bibinfo {volume} {77}},\ \bibinfo {pages} {041302} (\bibinfo {year} {2008})}\BibitemShut {NoStop}%
\bibitem [{\citenamefont {Hill}\ and\ \citenamefont {Paz}(2010)}]{Hill:2010yb}%
  \BibitemOpen
  \bibfield  {author} {\bibinfo {author} {\bibfnamefont {R.~J.}\ \bibnamefont {Hill}}\ and\ \bibinfo {author} {\bibfnamefont {G.}~\bibnamefont {Paz}},\ }\href {\doibase 10.1103/PhysRevD.82.113005} {\bibfield  {journal} {\bibinfo  {journal} {Phys. Rev. D}\ }\textbf {\bibinfo {volume} {82}},\ \bibinfo {pages} {113005} (\bibinfo {year} {2010})}\BibitemShut {NoStop}%
\bibitem [{\citenamefont {Lee}\ \emph {et~al.}(2015)\citenamefont {Lee}, \citenamefont {Arrington},\ and\ \citenamefont {Hill}}]{Lee:2015jqa}%
  \BibitemOpen
  \bibfield  {author} {\bibinfo {author} {\bibfnamefont {G.}~\bibnamefont {Lee}}, \bibinfo {author} {\bibfnamefont {J.~R.}\ \bibnamefont {Arrington}}, \ and\ \bibinfo {author} {\bibfnamefont {R.~J.}\ \bibnamefont {Hill}},\ }\href {\doibase 10.1103/PhysRevD.92.013013} {\bibfield  {journal} {\bibinfo  {journal} {Phys. Rev. D}\ }\textbf {\bibinfo {volume} {92}},\ \bibinfo {pages} {013013} (\bibinfo {year} {2015})}\BibitemShut {NoStop}%
\end{thebibliography}%
\end{document}